\documentclass[fleqn,10pt]{wlscirep}
\usepackage{graphicx}
\usepackage{epstopdf}
\usepackage[utf8]{inputenc}
\usepackage[T1]{fontenc}
\usepackage{bm}
\graphicspath{{Images/}}
\usepackage{multicol}
\usepackage{float}
\usepackage{comment}
\usepackage{amsmath}
\usepackage[misc]{ifsym}

\title{A universal DNA computing model for solving \textbf{NP}-hard subset problems}

\author[1]{Enqiang Zhu}
\author[1]{Xianhang Luo}
\author[2 $^{\textrm{\Letter}}$]{Chanjuan Liu}
\author[1 $^{\textrm{\Letter}}$]{Xiaolong Shi}
\author[3 $^{\textrm{\Letter}}$]{Jin Xu}
\affil[1]{Institute of Computing Science and Technology, Guangzhou University, Guangzhou 510006, China}
\affil[2]{School of Computer Science and Technology, Dalian University of Technology, Dalian 116024, China}
\affil[3]{School of Electronics Engineering and Computer Science, Peking University, Beijing 100871, China}

\affil[$^{\textrm{\Letter}}$]{Corresponding authors: chanjuanliu@dlut.edu.cn; xlshi@gzhu.edu.cn; jxu@pku.edu.cn}

\begin{abstract}
DNA computing, a nontraditional computing mechanism, provides a feasible and effective method for solving \textbf{NP}-hard problems because of the vast parallelism and high-density storage of DNA molecules. Although DNA computing has been exploited to solve various intractable computational problems, such as the Hamiltonian path problem, SAT problem, and graph coloring problem,  there has been little discussion of designing universal DNA computing-based models, which can solve a class of problems.
In this paper, by leveraging the dynamic and enzyme-free properties of DNA strand displacement, we propose a universal model named \texttt{DCMSubset} for solving subset problems in graph theory. The model aims to find a minimum (or maximum) set satisfying given constraints. For each element $x$ involved in a given problem, \texttt{DCMSubset} uses an exclusive single-stranded DNA molecule  to model $x$ as well as a specific DNA complex to model the relationship between $x$ and other elements.
Based on the proposed model, we conducted simulation and biochemical
experiments on three kinds of subset problems, a minimum dominating set, maximum independent set, and minimum vertex cover. We observed that \texttt{DCMSubset} can also be used to solve the graph coloring problem. Moreover, we  extended \texttt{DCMSubset} to a model for solving the SAT problem.
The results of experiments showed the feasibility and university of the proposed method.
Our results highlighted the potential for DNA strand
displacement to act as a computation tool to solve \textbf{NP}-hard problems.

\vspace{0.1cm}
\end{abstract}

\begin{document}

\flushbottom
\maketitle

\begin{multicols}{2}
\section*{Introduction}
The existence of \textbf{NP}-hard problems  that cannot be solved in a polynomial time (unless \textbf{P}=\textbf{NP}) has inspired researchers to exploit new computing models.  DNA computing, as a novel computing model, leverages the Watson--Crick complementary pairing
principle and predictable double helical structure, and has been extensively studied by researchers from various areas, including mathematics \cite{r1,r2,r3}, computer science \cite{r4,r5}, and biology \cite{r6,r7}. Original work on DNA computing was undertaken in 1994 by Adelman \cite{r8}, who provided an inspiring theory for solving intractable computational problems with biotechnology, and experimentally verified the principle by a simple example of the Hamiltonian path problem. Adelman's work created a new computing method for solving combinatorial problems, which is called molecular parallelism by Reif \cite{r9}. Although scholars have expressed skepticism toward molecular computation owing to its capability to handle only simple problems \cite{r10}, this argument was quickly destroyed by Lipton \cite{r11}, who demonstrated the feasibility of solving the SAT problem by extending Adelman's method. Recent studies in \cite{r1,r12} have shown that DNA computing can solve the graph coloring problem effectively.

In 1998, to simplify the process of solution detection, Roweis et al. \cite{r13} proposed a sticker-based model for DNA computation, called the sticker model, involving neither enzymes nor PCR extension. More importantly, the original double-stranded structures can be recovered through memory strands after secondary annealing, which allows for the reuse of the DNA material. Based on the sticker model, in 2002, Zimmermann et al. \cite{r14} designed DNA algorithms to solve the counting version of a series of \textbf{NP}-hard problems, including $k$-cliques, independent $k$-sets, Hamiltonian paths, and Steiner trees. Also in 2002,  Braich et al. \cite{r15} solved a 20-variable instance of the 3-SAT problem based on the separation operation of the sticker model. 

\begin{figure}[H]
  \centering
  \includegraphics[width=9cm]{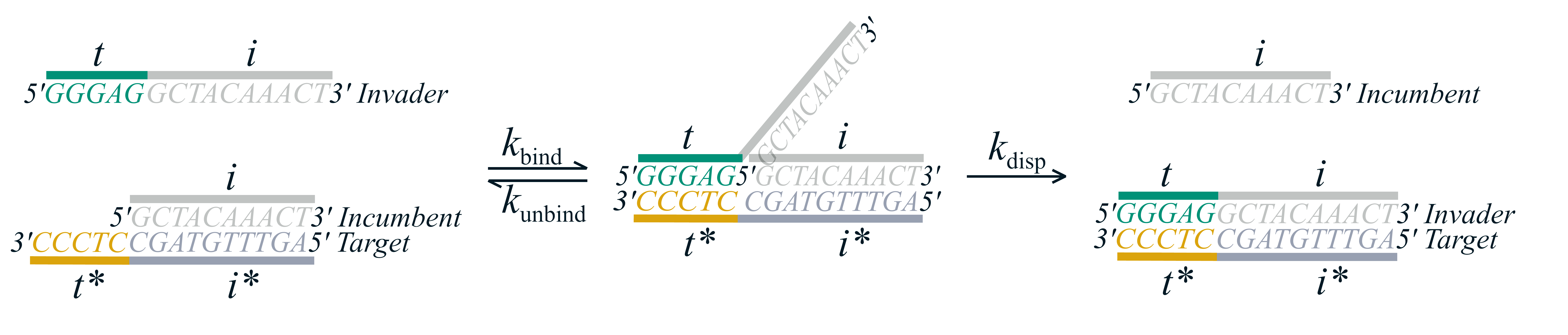}\\
  \caption{\small The principle of DNA strand displacement. Initially, an incumbent single-stranded $i$ is hybridized to a complementary domain $i^*$ of a target strand with an extra toehold domain $t^*$, forming a duplex. The invader consists of the incumbent single-stranded $i$ and the complementary toehold $t$ of $t^*$. The displacement starts with the action that the toehold $t$ of the invader binds to $t^*$  and progresses through a branch migration process of replacing the incumbent $i$ by the invader and creating a more stable double-stranded complex. The overall reaction rate is strongly dependent on toehold stability  \cite{r17,r18}.
  }\label{fig:Fig 1}
\end{figure}

Based on the sticker model, in 2000, Yurke \cite{r16} proposed the DNA strand displacement technology, which aims to construct nucleic acid systems with desired dynamic properties. A toehold-mediated strand displacement reaction is described as a molecular dynamic process of replacing a desired single-stranded DNA (called \emph{incumbent}) from a duplex (i.e., a double-stranded DNA with a sticky end) with an input single-stranded DNA (called \emph{invader}) that consists of the desired single-stranded DNA and the complementary strand of the sticky end (i.e., the toehold) to create a more stable complex (a double-stranded DNA) \cite{r17,r18}. Figure \ref{fig:Fig 1} illustrates  the principle of the strand displacement reaction. Compared with the traditional self-assembly models, DNA strand displacement can be completed spontaneously at room temperature. Moreover, the DNA strand displacement is enzyme-free, which provides with more flexibility in  constructing molecular circuits. Significant literature has been published on DNA strand displacement. In 2011, Qian and Winfree \cite{r5} proposed a reversible strand displacement logic gate, called a seesaw gate, based on which they designed a 4-bit square-root circuit  containing 130 DNA strands. The seesaw gate has been widely used as a simple building block for constructing large-scale circuits and neural networks \cite{r4}. In 2014, Machinek et al. \cite{r19} proposed a  method of creating mismatched base pairs to achieve the kinetic control of  strand displacement. In 2020, Wang et al. \cite{r20} designed DNA switching circuits based on DNA strand displacement for digital computing. Moreover, Liu et al. \cite{r17} proposed a DNA strand displacement circuit called the cross inhibitor, which is time-sensitive and allows interactive inhibition between two input signals. More recently, leveraging the property of programmable interactions between nucleic acid strands, Jung \cite{r6} expanded the capabilities of cell-free biosensors by designing DNA displacement interference circuits. Zhu et al. \cite{r21} first experimentally demonstrated that DNA strand displacement could be applied to encryption.

Several studies have utilized DNA strand displacement to solve intractable combinational optimization problems. In 2018, Tang et al. \cite{r22} applied DNA strand displacement to solve the 0-1 programming problem, where the model is built based on circular DNA. However, the method is impractical because the number of required species of fluorescence is equal to the number of variables. Therefore, in 2021, the same team \cite{r2} designed a chemical reaction network through three reaction modules (i.e., weighted, sum, and threshold) to solve 0-1 integer programming problems. The reliance on fluorescence was reduced significantly. Also Yang et al. \cite{r23} proposed a method to solve the SAT problem, which uses a specific origami structure to represent the solutions and  detects  the solutions by DNA strand displacement.

Although many models of DNA computation have been proposed to solve various types of problems,  many of these models were designed to solve only one type of problem.
As a result,  more energy and effort are required to design distinct models for different problems.  Several attempts have been made to reduce the operational complexity by simplifying the models, and great processes have been attained \cite{r24,r25}. However, different models require different environmental conditions (e.g., temperature, enzyme, and equipment), which hampers the compatibility between models, even though all these models are simplified extremely. This results in a high cost when using DNA computing to solve different problems, which hampers the popularization of molecular computers with powerful capability, like electronic computers.

To deal with the above challenges, researchers have  begun to design
universal models of DNA computation.
In 2019, Woods et al. \cite{r26} utilized tiles to construct an iterated boolean circuit, which can be used to simulate Turing machines \cite{r27}, universal boolean circuits \cite{r28}, and cellular automata \cite{r29}. In 2020, Wang et al. \cite{r20} designed a circuit for digital computation by combining DNA strand displacement and switching circuits, which can be applied to a variety of scenarios, such as molecular full adder \cite{r30} and the 4-bit square-root circuit \cite{r3,r31}.
In 2022, Xie et al. \cite {r32} designed a three-way junction-incorporated double hairpin unit by combining single-stranded gates \cite {r33} and exponential amplification reaction \cite {r34}, which can achieve multiple functions and applications, including a 4-to-2 encoder \cite {r35}, 1-to-2 demultiplexer \cite {r36}, 1-to-4 demultiplexer \cite {r37}, and multi-input OR gate.

However, regarding \textbf{NP}-hard problems, previous DNA computing models are still problem dependent, making the development of universal DNA computing models fascinating but challenging. Among \textbf{NP}-hard problems, there are many classic problems, such as the minimum dominating set, the maximum independent set, and the minimum vertex covering, which aim to find minimum or maximum subsets that satisfy some given restricted conditions. These problems belong to the subset problem in graph theory. This paper proposes a universal DNA computing model named \texttt{DCMSubset} for solving these problems. Given a vertex, \texttt{DCMSubset} utilizes  the Watson--Crick complementary pairing principle to model the relationship of the vertex  by a specific  DNA complex. The set of such DNA complexes corresponding to all vertices forms the computation gate circuit of \texttt{DCMSubset}.  Different elements (vertex or edge) are represented by different single-stranded sequences. Thus, each element is equipped with a separate detection gate that uses DNA strand displacement reactions to accurately detect the element. To demonstrate the feasibility of \texttt{DCMSubset}, we conducted simulation and biochemical experiments on the above three subset problems, the results of which
show the universality of \texttt{DCMSubset}. We observed that graph coloring could also be solved by \texttt{DCMSubset}, which is a direct application of the maximum independent set. Moreover, we extended \texttt{DCMSubset}  to a suitable model for solving the SAT problem. The results suggest the potential of \texttt{DCMSubset} to solve a wider range of \textbf{NP}-hard problems.

\setlength{\abovecaptionskip}{0.2cm}
\begin{figure*}[ht]
	\centering
	\includegraphics[width=17cm]{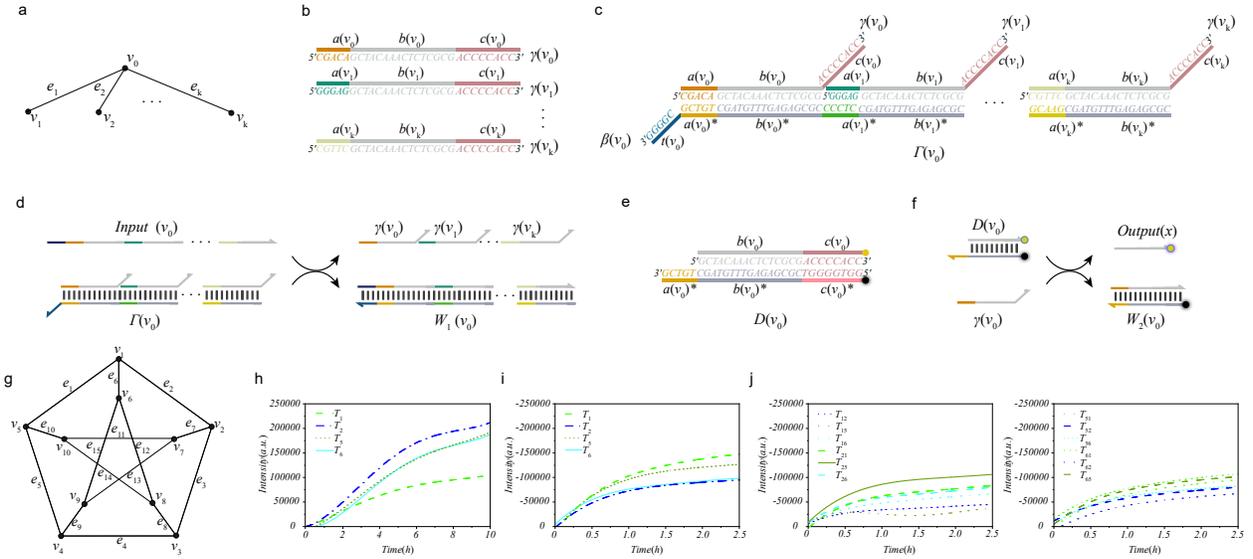}
	\caption{{\bf Principle of \texttt{DCMSubset}}. {\bf a} A star graph $S_k$. {\bf b} The structure of single strands $\gamma(v_i)$, $i=0,1,\ldots, k$, representing the vertices of $S_k$.  {\bf c} The structure of the computation gate $\Gamma(v_0)$, representing the adjacent relation between  $v_0$ and other vertices of $S_k$. {\bf d} The strand displacement reaction of $\Gamma(v_0)$ and the input single-stranded $Input(v_0)$, where $Input(v_0)$ is the complementary strand of $\beta(v_0)$.  {\bf e} The structure of detection gate $D(v_0)$ of the vertex $v_0$ of $S_k$.  {\bf f} The strand displacement reaction of $D(v_0)$ and  $\gamma(v_0)$.  {\bf g} The Peterson graph. {\bf h} Reaction kinetics of the experiment  showing the feasibility of \texttt{DCMSubset}.  {\bf i} Reaction kinetics of the first kind of possible leakage reactions.  {\bf j} Reaction kinetics of the second kind of possible leakage reactions. Where the curves $T_i$ (in Fig. \ref{fig:Fig 2} h and i) and $T_{ij}$ (in Fig. \ref{fig:Fig 2} j) for $i,j \in$ \{1,2,5,6\} and $i\neq j$ were  plotted by transferring the cycle value into the reaction time, indicating the change of the relative fluorescence unit (RFU) values in the FAM channel.}
	\label{fig:Fig 2}
\end{figure*}

\section*{Results}
{\bf The principle of \texttt{DCMSubset}.} Regarding subset problems, modeling elements and their relations using DNA molecules is fundamental to constructing universal models. This paper focuses on problems related to graphs, where a graph $G$ is often described as a 2-tuple ($V$, $E$) such that  $V (\neq \emptyset)$ represents the \emph{vertex set} of $G$,  and $E$ (a set of two-element subsets of $V$) represents the \emph{edge set} of $G$.  Two vertices are \emph{adjacent} if and only if they are the ends of an edge of $G$, while a vertex is \emph{incident} with an edge if and only if the vertex is an end of the edge. Given a vertex $u$, the set $N(u)$ of vertices adjacent to $u$ is called the \emph{neighborhood} of $u$, and $N[u]=N(u)\cup \{u\}$ is called the \emph{close neighborhood} of $u$; similarly, the set $N_e(u)$ of edges incident to $u$ is called the \emph{edge-neighborhood} of $u$, and $N_e[u]=N_e(u)\cup \{u\}$ is  the \emph{close edge-neighborhood} of $u$.

The problem we consider in this paper, named  $p$-SUBSET[$\phi$], can be formally described as follows.

\newtheorem{prob}{\bf Problem}
 \begin{prob}\label{prob1}
 $p$-SUBSET[$\phi$]: Given a graph $G=(V,E)$ and a property $\phi$, the problem asks to find a subset $S$ of $V$ with the minimum (or maximum) cardinality such that $S$ satisfies $\phi$.
 \end{prob}

Below is an account of the design of \texttt{DCMSubset} for solving $p$-SUBSET[$\phi$] problems.

Based on $\phi$, we first determine the underlying set $U$ of the problem;  generally, $U=V$ or $U=V\cup E$. For each element $x_0 \in U$, we design a 29-nt single-stranded DNA (denoted by $\gamma(x_0)$) to represent it and use a  DNA complex (denoted by $\Gamma(x_0)$) to characterize a relation (that is relevant to $\phi$, e.g.,  the adjacent relation and the incident relation) between $x_0$ and other elements. $\gamma(x_0)$ consists of a 5-nt toehold domain $a(x_0)$, a 16-nt branch migration domain $b(x_0)$, and a 8-nt branch migration domain $c(x_0)$. $\Gamma(x_0)$ is obtained by binding  all $\gamma(x_i)$s ($i=0,1,\ldots, k$) to a long single-stranded (denoted by $\beta(x_0)$) that  consists of a 5-nt toehold $t(x_0)$ at its 3'-end and  the complementary strands $a(x_j)^*$ and $b(x_j)^*$ of $a(x_j)$ and $b(x_j)$, respectively,  for $j\in \{0,\ldots, k\}$, where $x_j$  is either $x_0$ or an element associated with $x_0$. All $\Gamma(x)$s for $x\in U$ build the computation gate circuit of \texttt{DCMSubset}. To illustrate this modeling method, we consider the star graph $S_k$, shown in Fig. \ref{fig:Fig 2} a, where the vertex set $V=\{v_i|i=0,1,\ldots,k\}$ and the edge set $E=\{e_i=v_0v_i|i=1,2,\ldots,k\}$. We, for example, consider the adjacent relation and $U=V$. For $i=0,1,\ldots, k$,  $v_i$ is modeled as a  29-nt single-stranded $\gamma(v_i)$ (see Fig. \ref{fig:Fig 2} b)  and the close neighborhood  of $v_i$   is modeled as  a duplex $\Gamma(v_i)$ (see Fig. \ref{fig:Fig 2} c for an illustration of $\Gamma(v_0)$).

Each computation gate corresponding to an element $x\in U$ receives an exclusive input signal strand $Input(x)$ (the complementary strand of $\beta(x)$) as the invader to displace the incumbent strands, that is, $\gamma(x)$ and $\gamma(x_i)$ for $i=1,\ldots,k$, where  $x_i$ is an element associated with $x$. See Fig. \ref{fig:Fig 2} d for the strand displacement reaction of $\Gamma(v_0)$ and $Input(v_0)$, which releases $\gamma(v_i)$, $i=0,1,\ldots, k$ and a stable double strand ($W_1(v_0)$). The computation gate mechanism can be expressed via
the following reaction:
$$
\Gamma(x)+Input(x) \xrightarrow {k_1} \gamma(x)+\gamma(x_1)+\ldots+\gamma(x_k)+W_1(x)~~(1)
$$

To detecte the released single strands by the computation gate circuit, we need a detection gate for each element to recognize the released single strands. For each element $x$, the detection gate of $x$, denoted by $D(x)$, is a duplex that binds an incumbent strand consisting of $b(x)$ and $c(x)$ to an target strand consisting of $a(x)^*, b(x)^*$ and $c(x)^*$. Here, the incumbent strand carries a fluorophore at its  3'-end, and  the target strand carries a quencher at its  5'-end. Because  the quencher is close to the fluorophore, the fluorescence is inhibited in $D(x)$.  When the invader $\gamma(x)$ exists, its toehold $a(x)$ binds to the domain $a(x)^*$ of $D(x)$, and  branch migration
 moves gradually to the domain $c(x)^*$, which releases
a single-stranded $Output(x)$ with a fluorescence signal for monitoring and a stable beacon-labeled double-stranded $W_2(x)$. See Fig. \ref{fig:Fig 2} e and f for the structure of $D(v_0)$ and the strand displacement reaction between $D(v_0)$ and $\gamma(v_0)$, respectively.  All $D(x)$s build the detection gate circuit of \texttt{DCMSubset} for $x\in U$. The detection gate mechanism can be expressed via
the following reaction:
$$
~~~~~~~~~~~~~D(x)+\gamma(x) \xrightarrow {k_2} Output(x)+W_2(x)~~~~~~~~~~~~~~~~(2)
$$

\texttt{DCMSubset} consists of computation gate circuit and the detection gate circuit, which work as follows.

Regarding a $p$-SUBSET[$\phi$] problem $\mathcal{P}$ on the underlying set $U=\{v_1,v_2,\ldots, v_n\}$,  for the purpose of finding a minimum (or maximum) subset satisfying  $\phi$, we have to traverse over all the subsets of $U$. To determine whether a subset $U'=\{u_1,u_2,\ldots, u_{\ell}\} \subseteq U$ satisfies $\phi$, we first add
the computation gates $\Gamma$($u_{1}$), $\Gamma$($u_{2}$),\ldots, $\Gamma$($u_{\ell}$) into a tube $T$ and then divide $T$ equally into $n$ tubes $T_{1}$, $T_{2}$, \ldots, $T_{n}$. Second, we add the detection gates $D$($u_{i}$) into the $i$th tube $T_i$ for $i=1,2,\ldots, n$. Third, we add all $Input(u_i)$, $i=1,2,\ldots,\ell$, to each  $T_j$ for $j\in \{1,2,\ldots,n\}$ and detect the fluorescence signals. It should be noted that
computation gates should be designed  in compliance with the property $\phi$; that is,  the design should ensure that $U'$ is a solution of  $\mathcal{P}$ if and only if the combination of fluorescence signals in all $T_i$ ($i=1,2,\ldots,\ell$) is a certificate of the property $\phi$ (e.g., the fluorescence signals in all tubes are detected or none of the tubes are detected).

{\bf Feasibility of \texttt{DCMSubset}.}
To verify the feasibility of DCMSubset, we conducted a biochemical experiment for detecting the close neighborhood $N[v_1]=\{v_1, v_{2}, v_{5}, v_{6}\}$ of the vertex $v_1$ in the Peterson graph \textbf{P} (as shown in Fig. \ref{fig:Fig 2} g).
We first designed the input strands $Input(v_1)$, computation gate $\Gamma$($v_{1}$), and detection gates $D(v_{1})$, $D(v_{2})$, $D(v_{5})$, and $D(v_{6})$; see Supplementary Table S1 and S2 for the details of these strands. To verify that $\Gamma$($v_{1}$) can release single strands $\gamma(v_1), \gamma(v_2), \gamma(v_5)$ and  $\gamma(v_6)$, we found it sufficient to prepare four tubes (say $T_1, T_2, T_5,$ and $T_6$) such that $T_i$ contains the mixture of  $\Gamma(v_{1})$ and $D(v_i)$ for $i=1,2,5,6$.
For each $T_i$, we first added $D(v_i)$  (10 $\mu$$l$ $\times$ 1 $\mu$$M$) and the computation gates $\Gamma(v_1)$ (10 $\mu$$l$ $\times$ 2 $\mu$$M$); then, we added $Input(v_1)$ (10 $\mu$$l$ $\times$ 1 $\mu$$M$); finally, all $T_i$ were put into a fluorescence quantification PCR machine for detecting the fluorescence output.  The experimental results are shown in Fig. \ref{fig:Fig 2} h, from which we can see that the fluorescence signals in all these tubes are captured.

In addition, we can observe that the strand displacement reaction for  $D(v_i)$ (Eq. (2)) is triggered by the toehold domain $a(v_i)^*$ and proceeds from the branch migration domains   $b(v_i)^*$ and  $c(v_i)^*$; the input strand $Input(v_i)$ for $\Gamma(v_i)$ (Eq. (1)) contains $a(v_i)$ and $b(v_i)$; $\gamma(v_i)$ contains $a(v_i)$, $b(v_i)$, and $c(v_i)$. Therefore,  two kinds of experiments were carried out to analyze the possible leakage reactions: (1) the reactions of $Input(v_1)$ and $D(v_j)$ for $j=1,2,5,6$; and (2) the reactions of $\gamma(v_j)$ and $D(v_k)$ for $j,k \in \{1,2,5,6\}$ and $j\neq k$.

Regarding the first kind of experiment, we prepared four tubes $T_j$ ($j$ $\in$ $\{1,2,5,6\}$). For each tube $T_j$, we added the mixture of $D(v_j)$ (15 $\mu$$l$ $\times$ 0.67 $\mu$$M$) and  $Input(v_1)$ (15 $\mu$$l$ $\times$ 0.67 $\mu$$M$). Regarding the second experiment, we prepared 12 tubes denoted by $T_{jk}$ ($j,k \in \{1,2,5,6\}$ and $j\neq k$), to which we added $\gamma(v_i)$ (15 $\mu$$l$ $\times$ 1.33 $\mu$$M$) and $D(v_j)$ (15 $\mu$$l$ $\times$ 0.67 $\mu$$M$). To  monitor the fluorescence signals,  we put all the tubes $T_i$ and $T_{ij}$ into the fluorescence quantification PCR machine. It can be seen that  no fluorescence signal is detected  in all these tubes (see Fig. \ref{fig:Fig 2} i and j). This indicates  that the possible leakage reactions in  \texttt{DCMSubset} do not affect the accuracy of the detection gate.

Below is the applications of \texttt{DCMSubset} to three important subset problems in graph theory: the minimum dominating set, the maximum independent set, and the minimum vertex cover. All graphs considered in the paper contain no loop or parallel edge, i.e., simple graphs.

\setlength{\abovecaptionskip}{0.2cm}
\begin{figure*}[t]
	\centering
	\includegraphics[width=16cm,height=11cm]{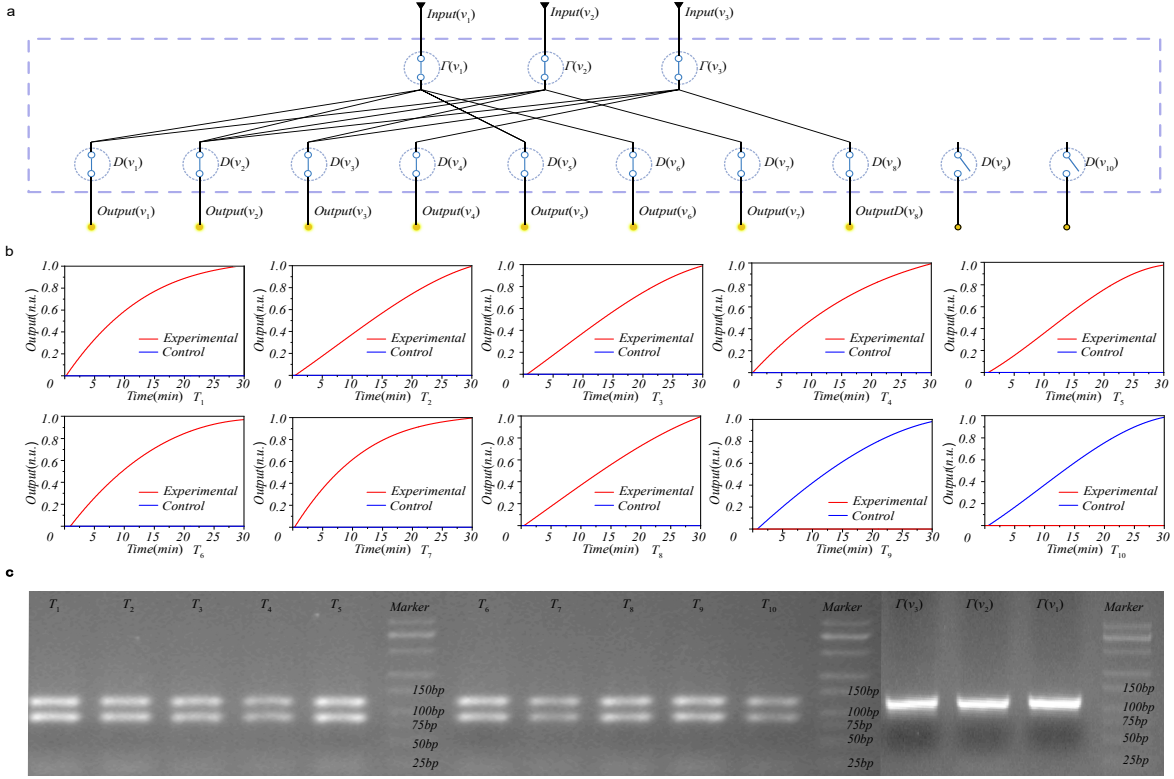}
	\caption{{\bf Experiments for determining whether  \{$v_{1}$,$v_{2}$,$v_{3}$\} is a DS of $\mathbf{P}$}. {\bf a} Logic circuit implementation of the first group of experiments. {\bf b} Reaction kinetics of the first group of experiments, which corresponds to the input \{$v_{1}$,$v_{2}$,$v_{3}$\}. The curve was plotted by transferring the cycle value to the reaction time. The outputs were normalized to RFU values in the FAM channel with the highest signals. {\bf c} Agarose gel electrophoresis results, where for $i=1,2,\dots,10$, $T_{i}$ contains input strands ($Input(v_{1})$, $Input(v_{2})$, and $Input(v_{3})$), computation gates ($\Gamma(v_{1}$), $\Gamma(v_{2})$, and $\Gamma(v_{3})$), and the detection gate $D(v_{i})$.} \label{fig:Fig 3}
\end{figure*}

\vspace{0.1cm}
{\bf Application to minimum dominating set.} Given a graph $G=(V,E)$, a \emph{dominating set} (DS) of $G$ is a subset $S$ of $V$ such that every vertex in $V\setminus S$ is adjacent to a vertex in $S$.
\begin{prob}
The \emph{minimum dominating set} (MDS) problem requires finding a DS of the minimum cardinality.
\end{prob}

We used \texttt{DCMSubset} to find an MDS of the Peterson graph $\mathbf{P}$ shown in Fig. \ref{fig:Fig 2} g. In order to reduce costs, we first dealt with the solution space by some preprocessing with a theoretical guarantee. We needed the following two results, where Proposition \ref{pro:1} follows directly from the definition of DS, and Proposition \ref{pro:2} was obtained by Bruce \cite{r38} in 1996.

\newtheorem{pro}{\bf Proposition}
 \begin{pro}\label{pro:1}
 If $S$ is a DS of a graph $G=(V,E)$, then $|N[S]|\geq |V|$, where $N[S]=\cup_{v\in S} N[v]$.
\end{pro}

\begin{pro}\label{pro:2}
Let $S$ be an MDS of a graph such that every vertex is adjacent to at least three vertices. Then, $|S|\leq \frac{3 |V|}{8}$.
\end{pro}

As $|N[v]|=4$ for any vertex of $\mathbf{P}$,  it has  $|N[S]|\leq 8$  for any $S\subset V$ with $|S|=2$. Thus, according to Proposition \ref{pro:1}, any dominating set of $\mathbf{P}$ contains at least three vertices. Moreover, by Proposition \ref{pro:2}, we know that any MDS of $\mathbf{P}$ contains exactly three vertices. Therefore, the solution space $\Omega$ is the set of all 3-subsets with cardinality ${10 \choose 3}=120$; that is, $\Omega=\{\{v_{1},v_{2},v_{3}\}, \{v_{1},v_{2},v_{4}\}, \ldots, \{v_{8},v_{9},v_{10}\}\}$. For description convenience, we used $S_i$, $i=1,2,\ldots, 120$ to denote the 120 candidate solutions in $\Omega$.

We now focus on finding an MDS of $\mathbf{P}$ using \texttt{DCMSubset}. As previously mentioned, for $i=1,2,\ldots, 10$, $v_i$ is modeled as a single-stranded $\gamma(v_i)$ that consists of three parts: $a(v_i)$, $b(v_i)$, and $c(v_i)$. The adjacent relation of $v_i$ is modeled as a duplex $\Gamma(v_i)$ (i.e., the computation gate of $v_i$), which is hybridized by $\gamma(x)$ for all $x \in N[v_i]$ and a long single-stranded $\beta(v_i)$, where $\beta(v_i)$ consists of a 5-nt toehold $t(v_i)$ and the complementary strands of $a(x)$ and $b(x)$. The detection gate $D(v_i)$ is a duplex that is hybridized by a single strand consisting of $b(v_i)$ and $c(v_i)$ and a single strand consisting of the complementary strands of  $a(v_i)^*$, $b(v_i)^*$, and $c(v_i)^*$. See Supplementary Tables S1 and S2 for the details of these strands.

The experiments are described as follows. We prepared 120 groups of experiments for testing the solution space, where each group consists of 10 tubes. The $i$th group of experiments was used to determine whether  $S_i$ is an MDS of $\mathbf{P}$. We use $T_{ij}$ to denote the $j$th tube in the $i$th group, where $i=1,2,\ldots, 120, j=1,2,\ldots, 10$.  We first added $D(v_j)$ (12.5 $\mu$l $\times$ 1.1 $\mu$M)  and the three computation gates $\Gamma(x), \Gamma(y),$ and  $\Gamma(z)$ (each 4 $\mu$l $\times$ 5.5 $\mu$M) corresponding to the three vertices in $S_i=\{x,y,z\}$
to each tube $T_{ij}$. Then, we added $Input(x), Input(y)$, and $Input(z)$ (each 2.5 $\mu$l $\times$  5.5 $\mu$M) to $T_{ij}$. Finally, all tubes $T_{ij}$ were put into the fluorescence quantification PCR machine for detecting the fluorescence output. See Fig. \ref{fig:Fig 3} a for the diagram of the first group of experiments, which corresponds the candidate solution $S_1$ = \{$v_1$, $v_2$, $v_3$\}.

According to the above discussion,  a vertex $s\in N[s']$ for some $s'\in $ $\{x,y,z\}$ implies that $s$ must be released by the displacement reactions of $\Gamma(s')$ and $Input(s')$. Therefore,

\newtheorem{thm}{\bf Theorem}
 \begin{thm}\label{thm-1}
 The $i$th candidate solution $S_i \in \Omega$ is a DS of $\mathbf{P}$ if and only if  the fluorescence signal in $T_{ij}$ is detected for all $j\in \{1,2,\ldots, 10\}$.
\end{thm}

Here, we emphasize that we stopped our experiments as soon as a solution was found, and all groups of experiments were conducted similarly (if all the 120 groups of experiments were implemented, then all of the MDS of $\mathbf{P}$ could be found). Thus,  only  12 groups of experiments were implemented,   corresponding to the inputs (candidate solutions)  $S_1=\{v_1,v_2,v_3\}$, $S_2=\{v_1,v_2,v_4\}$, $S_3=\{v_1,v_2,v_5\}$, $S_4=\{v_1,v_2,v_6\}$,
$S_5=\{v_1,v_2,v_7\}$, $S_6=\{v_1,v_2,v_8\}$, $S_7=\{v_1,v_2,v_9\}$, $S_8=\{v_1,v_2,v_{10}\}$, $S_9=\{v_1,v_3,v_4\}$, $S_{10}=\{v_1,v_3,v_5\}$,
$S_{11}=\{v_1,v_3,v_6\}$, and $S_{12}=\{v_1,v_3,v_7\}$. By Theorem \ref{thm-1}, $S_i$, $i=1,2,\ldots,12$ is a DS if all of the 10 tubes of the $i$th group of experiments output fluorescence signals. Figure \ref{fig:Fig 3} b shows the result of fluorescence  detection of the first group of experiments, in which the ninth and tenth tubes do not yield a fluorescence signal, which implies that $S_1$ is not a DS of $\mathbf{P}$. The results of fluorescence  detection of the other 11 groups of experiments are shown in Supplementary Fig. S1--S3, from which we see that $S_{12}$ is a DS of $\mathbf{P}$ and also an MDS of $\mathbf{P}$.

To test that the experiments can produce the desired outcomes (double-stranded $W_1$ in Eq. (1)), we separately carried out agarose gel electrophoresis experiments for pure $\Gamma(x)$  and  the mixture of $\Gamma(x), Input(x),$ and $D(x)$. Figure \ref{fig:Fig 3} c shows the experimental results corresponding to the three inputs $Input(v_1), Input(v_2)$, and $Input(v_3)$ (the results of the other 11 groups of experiments are shown in Supplementary Fig. S3), from which  we see that a new band between 75 bp and 100 bp is produced in the mixture solution,  while no new band is produced in the pure $\Gamma(x)$ solution.  This result is consistent with the design of \texttt{DCMSubset} because the double-stranded structure $W_1(x)$ formed by the strand displacement reaction of $\Gamma(x)$ and $Input(x)$ (Eq. (1)) has the length 89 bp.

\setlength{\abovecaptionskip}{0.2cm}
\begin{figure*}[t]
	\centering
	\includegraphics[width=16cm,height=6cm]{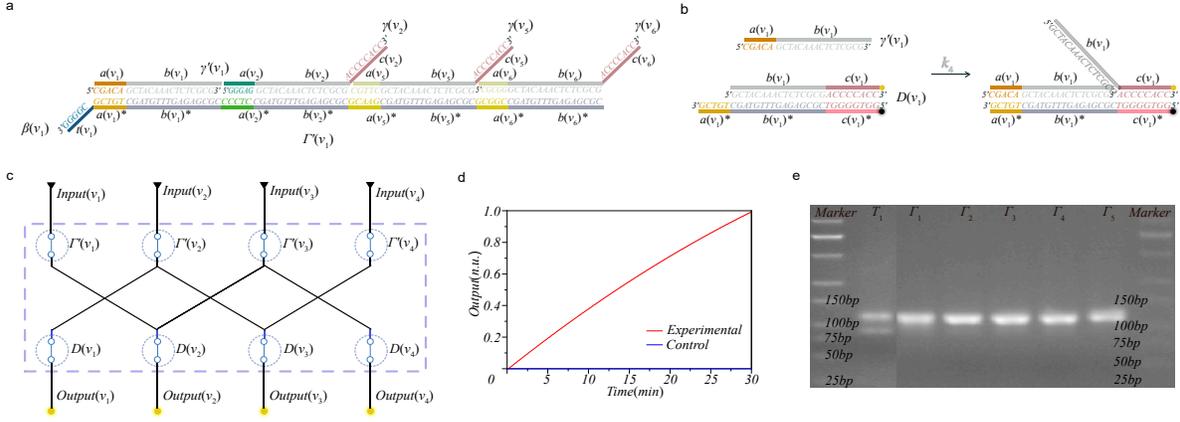}
	\caption{{\bf The first group of experiments that determine whether  \{$v_{1}$,$v_{2}$,$v_{3}$,$v_{4}$\} is an IS of $\mathbf{P}$.} {\bf a} The structure of computation gate $\Gamma'(v_1)$. {\bf b} The strand displacement reaction of $\gamma'(v_1)$ and the detection gate $D(v_1)$, where the incumbent strand of $D(v_1)$ (consists of domains $b(v_1)$ and $c(v_1)$) cannot be released. {\bf c} Logic circuit implementation.
 {\bf d} Reaction kinetics of the experiments, where the curve was plotted by transferring the cycle value into the reaction time, and the outputs were normalized to the RFU values in the FAM channel with the highest signals. {\bf e} Agarose gel electrophoresis results, where $T_1$ contains input strands ($Input(v_{1})$,$Input(v_{2})$,$Input(v_{3})$, and $Input(v_{4})$), computation gates ($\Gamma'(v_{1}$),$\Gamma'(v_{2})$,$\Gamma'(v_{3})$ and $\Gamma'(v_{4})$), and detection gates ($D(v_{1}$),$D(v_{2})$,$D(v_{3})$,$D(v_{4})$).}
	\label{fig:Fig 4}
\end{figure*}

\vspace{0.1cm}
{\bf Application to maximum independent set.} Given a graph $G=(V,E)$, an \emph{independent set} (IS) of $G$ is a subset $S\subseteq V$ such that any two vertices in $S$ are not adjacent.
\begin{prob}
The \emph{maximum independent set} (MIS) problem requires finding an IS of the maximum cardinality.
\end{prob}
We used \text{DCMSubset} to find an \text{MIS} of Peterson graph $\mathbf{P}$ (see Fig. \ref{fig:Fig 2} h). Analogously, we first reduced  the solution space by some preprocessing with theoretical guarantee. We utilized  the following two results: Proposition \ref{pro-3} follows directly by the  pigeonhole principle and the fact that the set of vertices with the same color is an independent set; Proposition \ref{pro-4} was obtained by Lov{\'{a}}sz in 1979 \cite{r39}.

\begin{pro}\label{pro-3}
	Let $I$ be an IS of an $n$-vertex graph $G$. Then, $|I|$ $\geq$ $\lfloor\frac{n}{\chi(G)}\rfloor$, where $\chi(G)$ is the chromatic number of $G$, that is, the minimum number of colors assigned to the vertices of $G$ such that no two adjacent vertices receive the same color.
\end{pro}

\begin{pro}\label{pro-4}
	Let $I$ be an MIS of an $n$-vertex graph $G$.  Then, $|I|$ $\leq$ $\lceil\frac{-n\lambda_{min}(A)}{\lambda_{max}(A)-\lambda_{min}(A)}\rceil$, where $A$ is the adjacency matrix of $G$, and $\lambda_{max}(A)$ and $\lambda_{min}(A)$ are the maximum and minimum eigenvalues of $A$, respectively.
\end{pro}

As the chromatic number of a graph that is neither isomorphic to $K_4$ (the complete graph on four vertices) nor an odd cycle is at most the maximum degree \cite{r40}, it follows that $\chi(\mathbf{P})\leq 3$, so any MIS of $\mathbf{P}$ contains at least four vertices, according to Proposition \ref{pro-3}. Additionally, the maximum and minimum eigenvalues of the adjacency matrix of $\mathbf{P}$ are 3 and -2, respectively, which implies that an MIS of $\mathbf{P}$ contains at most four vertices, according to Proposition \ref{pro-4}. Therefore, an MIS of $\mathbf{P}$ contains exactly four vertices, and the solution space $\Omega$ is the set of all 4-subsets with the cardinality $\binom{10}{4}$=210; that is, $\Omega$ = \{\{$v_1$, $v_2$, $v_3$, $v_4$\},\{$v_1$, $v_2$, $v_3$, $v_5$\},\ldots,\{$v_7$, $v_8$, $v_9$, $v_{10}$\}\}. We use $I_k$, $k = 1, 2, \ldots, 210$ to denote the 210 candidate solutions in $\Omega$.

We now focus on  the experiment of finding an MIS of $\mathbf{P}$ base on \text{DCMSubset}. All key components  were designed in a manner similar to the MDS problem except for a slight change to the computation gate $\Gamma(v_i), i=1,2,\ldots, 10$. Here, each $\Gamma(v_i)$ was modified by just deleting the $c(v_i)$ component of $\gamma(v_i)$ at the 5'-end, and the other components were unchanged. We denote $\Gamma'(v_i)$ as the new computation gate; see Fig. \ref{fig:Fig 4} a for $\Gamma'(v_0)$. The experiments were conducted as follows. We prepared 210 groups of experiments for testing the solution space, where each group used one tube. The $k$th group of experiments (denoted by $T_{k}$)  aimed to determine whether  $I_k$ is an MIS of $\mathbf{P}$. Suppose that $I_k$ = $\{x,y,z,w\}$.
We first added the four detection gates $D(x)$, $D(y)$, $D(z)$, $D(w)$ (each 5 $\mu$$l$ $\times$ 0.916 $\mu$$M$) and the four computation gates $\Gamma'(x)$, $\Gamma'(y)$, $\Gamma'(z)$ and $\Gamma'(w)$ (each 5 $\mu$$l$ $\times$ 1.832 $\mu$$M$)  to $T_k$. Then, we added $Input(x)$, $Input(y)$, $Input(z)$ and $Input(w)$ (each 5 $\mu$$l$ $\times$ 0.916 $\mu$$M$) to $T_k$. Finally, $T_k$ was put into a fluorescence quantification PCR machine for detecting the fluorescence output. See Fig. \ref{fig:Fig 4} c for the circuit diagram of the first group of experiments, which corresponds to the solutions $I_1$ = \{$v_1$, $v_2$, $v_3$, $v_4$\}.

It is clear that if an arbitrary vertex $s\in \{x,y,z,w\}$ is adjacent to one of $x,y,z,$ and $w$, then $\gamma$($s$) is released by  $\Gamma'$($s'$) for some $s'\in \{x,y,z,w\}$ and then detected by $D(s)$; moreover, if $\{x,y,z,w\}$ is an IS, then no $\gamma$($s$) for $s\in \{x,y,z,w\}$ is released by  $\Gamma'$($s'$) for any $s'\in \{x,y,z,w\}$. We can observe that $\Gamma^{\prime}(s)$ and $Input(s)$ can only produce a partial strand of $\gamma$($s$) that consists of only $a(s)$ and $b(s)$, denoted by $\gamma^{\prime}(s)$. Also, $\gamma^{\prime}(s)$ and $D(s)$ do not produce fluorescence signals (see Fig. \ref{fig:Fig 4} b).
Therefore, the following result holds.

\begin{thm}\label{thm---2}
	The $k$th candidate solution $I_k$ $\in$ $\Omega$ is an IS of $\mathbf{P}$ if and only if no fluorescence signal is detected in  $T_k$.
\end{thm}

Also, we stopped our experiments as soon as a solution was found. Thus, we only conducted 49 groups of experiments, corresponding to the candidate solutions  \{$v_1$,$v_2$,$v_3$,$v_4$\},  \{$v_1$,$v_2$,$v_3$,$v_5$\}, \{$v_1$,$v_2$,$v_3$,$v_6$\}, \{$v_1$,$v_2$,$v_3$,$v_7$\}, \{$v_1$,$v_2$,$v_3$,$v_8$\}, \{$v_1$,$v_2$,$v_3$,$v_9$\}, \{$v_1$,$v_2$,$v_3$,$v_{10}$\}, \{$v_1$,$v_2$,$v_4$,$v_5$\}, \{$v_1$,$v_2$,$v_4$,$v_6$\}, \{$v_1$,$v_2$,$v_4$,$v_7$\}, \{$v_1$,$v_2$,$v_4$,$v_8$\}, \{$v_1$,$v_2$,$v_4$,$v_9$\}, \{$v_1$,$v_2$,$v_4$,$v_{10}$\}, \{$v_1$,$v_2$,$v_5$,$v_6$\}, \{$v_1$,$v_2$,$v_5$,$v_7$\}, \{$v_1$,$v_2$,$v_5$,$v_8$\}, \{$v_1$,$v_2$,$v_5$,$v_9$\}, \{$v_1$,$v_2$,$v_5$,$v_{10}$\}, \{$v_1$,$v_2$,$v_6$,$v_7$\}, \{$v_1$,$v_2$,$v_6$,$v_8$\}, \{$v_1$,$v_2$,$v_6$,$v_9$\}, \{$v_1$,$v_2$,$v_6$,$v_{10}$\}, \{$v_1$,$v_2$,$v_7$,$v_8$\}, \{$v_1$,$v_2$,$v_7$,$v_9$\}, \{$v_1$,$v_2$,$v_7$,$v_{10}$\}, \{$v_1$,$v_2$,$v_8$,$v_9$\}, \{$v_1$,$v_2$,$v_8$,$v_{10}$\}, \{$v_1$,$v_2$,$v_9$,$v_{10}$\},
\{$v_1$,$v_3$,$v_4$,$v_{5}$\}, \{$v_1$,$v_3$,$v_4$,$v_{6}$\},
\{$v_1$,$v_3$,$v_4$,$v_{7}$\}, \{$v_1$,$v_3$,$v_4$,$v_{8}$\},
\{$v_1$,$v_3$,$v_4$,$v_{9}$\}, \{$v_1$,$v_3$,$v_4$,$v_{10}$\},
\{$v_1$,$v_3$,$v_5$,$v_{6}$\}, \{$v_1$,$v_3$,$v_5$,$v_{7}$\},
\{$v_1$,$v_3$,$v_5$,$v_{8}$\}, \{$v_1$,$v_3$,$v_5$,$v_{9}$\},
\{$v_1$,$v_3$,$v_5$,$v_{10}$\}, \{$v_1$,$v_3$,$v_6$,$v_{7}$\},
\{$v_1$,$v_3$,$v_6$,$v_{8}$\}, \{$v_1$,$v_3$,$v_6$,$v_{9}$\},
\{$v_1$,$v_3$,$v_6$,$v_{10}$\}, \{$v_1$,$v_3$,$v_7$,$v_{8}$\},
\{$v_1$,$v_3$,$v_7$,$v_{9}$\}, \{$v_1$,$v_3$,$v_7$,$v_{10}$\},
\{$v_1$,$v_3$,$v_8$,$v_{9}$\}, \{$v_1$,$v_3$,$v_8$,$v_{10}$\},
and \{$v_1$,$v_3$,$v_9$,$v_{10}$\}, where \{$v_1$,$v_3$,$v_9$,$v_{10}$\} is an MIS, while the others are not ISs of $\mathbf{P}$. Figure \ref{fig:Fig 4} d shows the result of fluorescence  detection of the first group of experiments, which implies that $S_1$ is not an IS of $P$. Figure \ref{fig:Fig 4} e shows the result of agarose gel electrophoresis of the first group of experiments. The results of the other 48 groups of experiments are shown in Supplementary Figs. S4--S5, from which we see that $S_{49}$ is an IS of $\mathbf{P}$ and an MIS of $\mathbf{P}$.

As a direct extension, we observe that the \texttt{DCMSubset} circuit can be used to solve the $k$-coloring problem, which requires us to determine whether there exists an assignment of $k$ colors to the vertices of a graph $G$ such that no two adjacent vertices receive the same color. As  such an assignment is a $k$-coloring of $G$ if and only if every set of vertices that receive  the same color is an IS of $G$, it is enough to enumerate all $k$-partitions of $V(G)$ (i.e., divide $V(G)$ into $k$ disjoint subsets), and for each  $k$-partition $V_1,V_2,\ldots, V_k$ (where $V_1\cup V_2 \cup \ldots\cup V_k=V(G)$ and $V_i\cap V_j=\emptyset$ for $i,j\in \{1,2,\ldots, k\}$ and $i\neq j$), we can determine whether $V_i$ (for $i=1,2,\ldots,k$) is an IS of $G$ using the \texttt{DCMSubset} circuit for the MIS problem. If $V_i$ is an IS for all $i\in \{1,2,\ldots,k\}$, then $V_1,V_2,\ldots, V_k$ corresponds to a $k$-coloring (vertices in $V_i$ is colored with $i$); otherwise, $V_1,V_2,\ldots, V_k$ does not correspond to a $k$-coloring, and we consider the next $k$-partition.

 Below we argue the applications of \texttt{DCMSubset} to the minimum vertex cover problem.  As the experiments were implemented in the same way as  the MDS problem, we only verified the correctness of the designed circuits by simulation experiments.

\vspace{0.1cm}
{\bf Application to minimum vertex coverage.}  Given a graph $G=(V,E)$, a \emph{vertex covering} (VC)  of $G$ is a subset $C\subseteq V$ such that each edge is incident with at least one vertex of $C$.
\begin{prob}
The \emph{minimum vertex covering} (MVC) problem requires us to find a VC of the minimum cardinality.
\end{prob}
\setlength{\abovecaptionskip}{0.2cm}
\begin{figure*}[t]
	\centering
	\includegraphics[width=16cm,height=3.5cm]{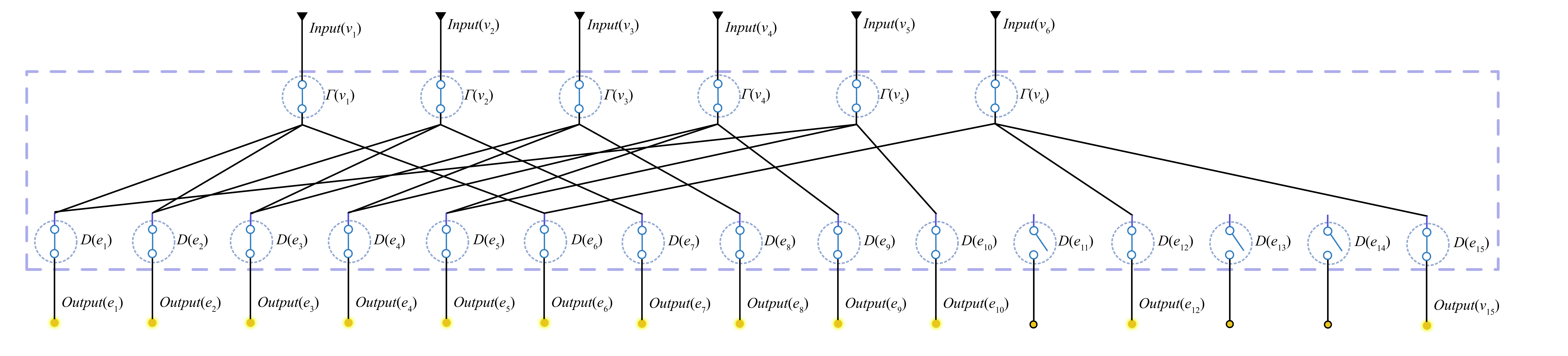}
	\caption{{\bf Logic circuit
implementation of determining whether or not \{$v_{1}$,$v_{2}$,$v_{3}$,$v_{4}$,$v_{5}$,$v_{6}\}$ is a VC of $\mathbf{P}$ based on \texttt{DCMSubset}. }}
	\label{fig:Fig 5}
\end{figure*}

Next, we constructed the \texttt{DCMSubset} circuit for finding all MVCs of the Peterson graph $\mathbf{P}$. Because an MVC involves both vertices and edges of a graph, we had to model both of these two types of elements. While all of the  key components of the circuit were designed in the same way as that for the MDS problem,  single-stranded $\gamma(e_i)$ and complex $D(e_i)$, for $i=1,2,\ldots, 15$ were added to model the edges of $\mathbf{P}$ and their detection gate. Moreover, each computation gate $\Gamma(v_i)$, $i=1,2,\ldots, 10$, indicating the incident relation, was hybridized by $\gamma(x)$ for $x\in \{v_i, e_{i_1}, e_{i_2}, e_{i_3}\}$ and a long single-stranded $\beta(v_i)$ consisting of  a
5-nt toehold $t(v_i)$ and the complementary domains of $a(x)$ and
$b(x)$. Here, $e_{i_1}$, $e_{i_2}$, and $e_{i_3}$ are the three edges that are incident with $v_i$. See Supplementary Fig. S8 for the illustration of $\Gamma(v_1)$.

For an $n$-vertex graph $G$, it is well known that $|I|+|C|=n$ (first observed by Gallai \cite{r41} in 1959), where $I$ and $C$ are an MIS and  MVC of $G$. Therefore, an MCV of $\mathbf{P}$ contains six vertices, and the solution space $\Omega$ is the set of all 6-subsets of $V(\mathbf{P})$ with the cardinality $10 \choose 6=210$. Let
$\Omega = \{C_i|i=1,2,\ldots,210\}$.

The simulation experiments were implemented as follows. We prepared 210 groups of experiments. For each $i=1,2,\ldots, 210$, the $i$th group of experiments
involves 15 specific experiments, for the purpose of determining whether $C_i$ is an MVC of $\mathbf{P}$. We use $T_{ij}$ to denote the $j$th experiment in $i$th group, where  $j=1,2,\ldots, 15$.  Let $C_i=\{v_{i_k}|k=1,2,\ldots,6\}$. For each group $T_{ij}$, we first added $D(e_j)$ ($10nM$), then the six computation gates $\Gamma(x_{i_k})$ (each $20nM$), and finally the six inputs $Input(x_{i_k})$(each $10nM$), where $k=1,2,\ldots, 6$.  To assess whether $C_i$ is a VC, we detected the concentration of the output signal  using DSD simulation tool.  See Fig. \ref{fig:Fig 5} for the circuit of the first group of experiments, which corresponds to the inputs $Input(v_{k})$ for $k=1,2,\ldots, 6$.

Through the above analysis, we obtained the following result.
{
		\begin{thm}\label{thm-3}
			$C_i \in \Omega$ is a VC of $\mathbf{P}$ if and only if the concentration of the output signal in $T_{ij}$ is detected for all $j\in \{1,2,\ldots, 15\}$.
\end{thm}}

We conducted 210 groups of simulation experiments, corresponding to the 210 candidate solutions  \{$v_1$,$v_2$,$v_3$,$v_4$,$v_5$,$v_6$\},  \{$v_1$,$v_2$,$v_3$,$v_4$,$v_5$,$v_7$\}, $\ldots$, \{$v_5$,$v_6$,$v_7$,$v_8$,$v_9$,$v_{10}$\}, where \{$v_{1}$,$v_{2}$,$v_{4}$,$v_{8}$,$v_{9}$,$v_{10}$\}, \{$v_{1}$,$v_{3}$,$v_{4}$,$v_{6}$,$v_{7}$,$v_{10}$\}, \{$v_{1}$,$v_{3}$,$v_{5}$,$v_{7}$,$v_{8}$,$v_{9}$\}, \{$v_{2}$,$v_{3}$,$v_{5}$,$v_{6}$,$v_{9}$,$v_{10}$\}, and \{$v_{2}$,$v_{4}$,$v_{5}$,$v_{6}$,$v_{7}$,$v_{8}$\} are MVCs. The simulation results are illustrated in Supplementary Figs. S11--S17.

\section*{Discussion}
We demonstrated that \texttt{DCMSubset} could be applied to  subset problems in graph theory. In contrast with previous DNA computing models, \texttt{DCMSubset} is more general.


{\bf Splitting computation gates.} We observed that when a vertex $x$ had a big neighborhood, the single-stranded $\beta(x)$ in the computation gate $\Gamma(x)$ was significantly longer. This may increase  the likelihood of error in the process of molecular assembly.  We attempted to solve this problem by a  splitting method with regard to computation gates. Consider the  computation gate $\Gamma(v_0)$ shown in Fig. \ref{fig:Fig 2} c, which is obtained by binding $\gamma(v_i)$,  $i=0,1,\ldots, k$ to a  single-stranded $\beta(v_0)$. It is clear that  the length of $\beta(v_0)$ increases with the increase of $k$. We controlled the length of $\beta(v_0)$ by splitting $\{v_0,v_1,\ldots, v_k\}$ into $\ell$ subsets with a similar cardinality. We took $\ell=3$ as an example to show the splitting method, and other cases could be dealt with similarly. First, construct a covering of $\{v_0,v_1,\ldots, v_k\}$ that consists of  three subsets $N_1=\{v_0,v_1,\dots,v_{r_1}\}, N_2=\{v_0, v_{r_1+1},v_{r_1+2},\dots,v_{r_2}\}$, and $N_3=\{v_0, v_{r_2+1},v_{r_1+r_2+2},\dots,v_{k}\}$, where $r_1=\lfloor\frac{k}{3}\rfloor$ and $r_2=r_1+\lfloor\frac{k-r_1}{2}\rfloor$. Then,  the computational gate $\Gamma(v_0)$ (see Supplementary Fig. S10 a) of $v_0$ is split into three components ${\Gamma(v_0)_1}$ (see Supplementary Fig. S10 b), ${\Gamma(v_0)_2}$ (see Supplementary Fig. S10 c), and ${\Gamma(v_0)_3}$ (see Supplementary Fig. S10 d), where for $i=1,2,3$, ${\Gamma(v_0)_i}$ is the duplex hybridized by all $\gamma(x)$ for $x\in N_i$  and a single-stranded $\beta_i(v_0)$ that consists of a 5-nt toehold $t$ at its 3'-end and the complementary
strands of $a(x)$ and $b(x)$ for $x\in N_i$. Here, a slight change to  ${\Gamma(v_0)_2}$ and ${\Gamma(v_0)_3}$ should be done by deleting the $c(v_0)$ domain from their $\gamma(v_0)$ strand (we use $\gamma^{\prime}(v_0)$ to denote the remainder of $\gamma(v_0)$ after deleting its $c(v_0)$ domain). Correspondingly,  the input single-stranded $Input(v_0)$ is  replaced by three short single strands $Input(v_0)_1$, $Input(v_0)_2$, and $Input(v_0)_3$ (see Supplementary Fig. S10 e), where $Input(v_0)_i$ is the complementary $\beta_i^*(v_0)$ of  $\beta_i(v_0)$, for $i=1,2,3$.  Because $\gamma^{\prime}(v_0)$ and $D(v_0)$ do not produce fluorescence signal (see the previous experiments for the MIS problem, as shown in Fig. \ref{fig:Fig 4} b), the strand displacement reaction for computation gate (Eq. (1)) can be replaced by  the following three reactions.
\vspace{0.1cm}
$
\noindent \mathrm{\Gamma}(v_0)_1+Input(v_0)_1\xrightarrow {k_3} \gamma(v_0)+\gamma(v_1)+\ldots+\gamma(v_{r_1})+{W}_1
$
$
\noindent\mathrm{\Gamma}(v_0)_2+Input(v_0)_2\xrightarrow {k_3} \gamma^{\prime}(v_0)+\gamma(v_{r_1+1})+\ldots+\gamma(v_{r_2})+{W}_2
$
$
\noindent\mathrm{\Gamma}(v_0)_3+Input(v_0)_3\xrightarrow {k_3} \gamma^{\prime}(v_0)+\gamma(v_{r_2+1})+\ldots+\gamma(v_k)+{W}_3
$

\vspace{0.1cm}
{\bf Application to SAT.} In the previous sections, we demonstrated that \texttt{DCMSubset} can be used to solve subset problems in graph theory. To further extend the scope of its applicability, we attempted to solve other problems by \texttt{DCMSubset}. A \texttt{DCMSubset}-based approach for the SAT problem is described as follows.
\begin{prob}
The SAT problem aims to determine whether  there is a truth assignment of 0 (`false') or 1 (`true')  to a set of boolean variables $\{x_1,\ldots, x_n\}$ that makes a conjunctive normal
form (CNF)  $F=C_1 \wedge C_2 \wedge \ldots \wedge C_m$  true, where $C_i=\ell_{i1}\vee \ell_{i2} \vee \ldots \vee \ell_{ik_i}$  for $i=1,2,\ldots,m$  and $\ell_{ij}$ ($j=1,2,\ldots, k_i$) is called a literal, which is either $x$ or the negative $\neg x$ of $x$ for some $x\in \{x_1,\ldots, x_n\}$.
\end{prob}

It is clear that $F$ has true value if and only if every clause $C_i$ has true value.
Our target is to design a DNA circuit based on the method of \texttt{DCMSubset} to determine the truth value of each clause $C_i$ for a given truth assignment to the boolean variables.

Unlike in the models for graph problems, for each $i\in \{1,2,\ldots,n\}$, we used single strands $\gamma^+(x_i)$ and $\gamma^-(x_i)$ (see Supplementary Fig. S6 a) to present $x_i$ and $\neg x_i$, respectively. Here, $\gamma^+(x_i)$ and $\gamma^-(x_i)$ were designed as follows: In the direction from 5' to 3', $\gamma^+(x_i)$ consists of a toehold domain $a(x_i)$ and two branch migration domains $b(x_i)$ and $c(x_i)$; $\gamma^-(x_i)$ consists of the complementary domains  $a(x_i)^*$, $b(x_i)^*$, and $c(x_i)^*$ of $a(x_i)$, $b(x_i)$, and $c(x_i)$. Observe that although the bases between $\gamma^+(x_i)$ and $\gamma^-(x_i)$ are complementary along the 5' to 3' direction, they are not hybridized into a double strand owing to the restriction of their directions.

The computation gate $\Gamma(C_i)$  was designed for each clause $C_i$, $i=1,2,\ldots, m$, obtained by binding all $\gamma(x)$ for $x\in C_i$ (here, $\gamma(x)$=$\gamma^-(x)$ if the corresponding literal is the negative form of $x$; otherwise, $\gamma(x)$=$\gamma^+(x)$) to a single-stranded $\beta(C_i)$, via the toehold domain $a(x)$ (or $a(x)^*$) and branch migration domain $b(x)$ (or $b(x)^*$). Note that $\beta(C_i)$ was designed different from the previous design; we reserved a toehold domain $t$ for every $x\in C_i$;  see Supplementary Fig. S6 c.

Now, for each truth assignment $\theta=\{\theta_1, \theta_2,\ldots, \theta_n\}$, where $\theta_i \in \{0,1\}$ represents the truth value of the variable $x_i$, we designed $n$ input strands $Input(x_i)$ (see Supplementary Fig. S6 b), $i=1,2,\ldots, n$ as follows: When $\theta_i=0$, $Input(x_i)$ is denoted by $Input^{-}(x_i)$, which (in the direction from 5' to 3') consists of the toehold domain $t^*$ and the complementary domains  $a(x_i)^*$ and $b(x_i)^*$ of $a(x_i)$ and $b(x_i)$, respectively; when $\theta_i=1$, $Input(x_i)$ is denoted by $Input^{+}(x_i)$, which (in the direction from 5' to 3') consists of the toehold domain $t^*$ and two branch migration domains $a(x_i)$ and $b(x_i)$. For the same reason as that for $\gamma^+(x_i)$ and $\gamma^-(x_i)$, $Input^+(x_i)$ and $Input^-(x_i)$ cannot hybridize into a double-stranded complex.

Each computation gate $\Gamma(C_i)$  receives $k_i$ input signal strands ${Input(x_{i_j})}$, $j=1,2,\dots,k_i$) as the invader to displace the incumbent strands $\gamma(x_{i_j})$, where $C_i=\ell_{i1}\vee \ell_{i2} \vee \ldots \vee \ell_{ik_i}$). If and only if ${Input(x_{i_j})}$ and  $\gamma(x_{i_j})$ have the same symbol (i.e., either $Input^+(x_{i_j})$ and $\gamma^+(x_{i_j})$, or $Input^-(x_{i_j})$ and $\gamma^-(x_{i_j})$), the incumbent strands  $\gamma(x_{i_j})$ can be replaced by the strand displacement reaction of  this group of inputs and $\Gamma(C_i)$.

The detection gate $D(\ell_{ij})$ was designed for each literal in $C_i$, $i=1,2,\ldots, m$, $j=1,2,\ldots, k_i$, which is obtained by binding the target strand (a complementary strand $\gamma(\ell_{ij})^*$ of  $\gamma(\ell_{ij})$) to a single-stranded output, via the branch migration domains $b(x)$ (or $b(x)^*$) and  $c(x)$ (or $c(x)^*$). When the invader $\gamma(\ell_{ij})$ exists, the strand displacement reaction of $\gamma(\ell_{ij})$ and $D(\ell_{ij})$ can occur and produce an output single. Supplementary Fig. S6 d illustrates the work flow diagram of \text{DCMSubset} on SAT. In Supplementary Fig. S6 c, we give a molecular implementation of the \text{DCMSubset} circuit for the SAT problem\label{key}.

For each truth assignment $\theta$, we carried out  $i$ $(i=1,2,\dots,m)$ groups of experiments to determine the truth value of each clause $C_i$. $T_i$ denotes the $i$th group of experiments. Regrading each group of experiments,  we added $\Gamma(C_i)$ ($10 nM$), $D(\ell_{ij})$ for $j=1,2,\ldots, k_i$ (each $10 nM$), and inputs corresponding to the truth assignment  (each $Input(x_i)$ $10nM$ and total $(10*n)nM$) to $T_i$. Then, we detected the content change of output signal by DSD simulation. According to the above analysis, the truth value of $C_i$ is true if and  only if  the output content increases in $T_i$. Therefore, $F$ is true under $\theta$ if and only if the output content increases in all these $m$ groups of experiments.  See the supplementary case analysis for an example of solving the SAT problem.

Based on our results, we posit that \texttt{DCMSubset} can also be used to design DNA circuits for other classes of \textbf{NP}-hard problems, such as the Hamiltonian problem, which we will attempt to realize in our future work.

\section*{Methods}
{\bf Materials and apparatuses.} The DNA oligonucleotides (oligos) used in this study were purchased from Sangon Biotech (Shanghai). All DNA oligos were purified by Sangon using HPLC. Agarose, 10$\times$TE Buffer, 50$\times$TAE Buffer, 4S Green Plus Nucleic Acid Stain, Sterilized ddH$_2$O, and DNA Marker A (25-500bp) were purchased from Sangon Biotech. The DNA loading buffer (6$\times$) was purchased from biosharp (Guangzhou). The concentration of DNA oligonucleotides was measured using a NanoPhotometer$^@$ N120 (Implen, Munich, Germany). All samples were annealed in a polymerase chain reaction thermal cycler (Thermo Fisher Scientific Inc., USA). The PCR analyses were carried out in QuantStudio 3, using the QuantStudio Design $\&$ Analysis Software v.1.5.1 (Applied Biosystems, MA, US). Agarose gel electrophoresis was carried out in BG-subMINI with BG-Power 600 (BayGene, Beijing). The DNA image on the gel was taken on AlphaImager HP (Protein Simple, San Jose, CA) gel documentation system.

Individual unlabeled DNA oligos were dissolved in 1$\times$TE buffer (nuclease free, pH 7.8$\sim$8.2) containing 12.5 mM Mg$^{2+}$ to a final concentration of 100 $\mu$M according to the information provided by the supplier, and stored at 4 $^{\circ}$C. Oligos labeled with dyes or quenchers were dissolved in Sterilized ddH$_2$O to a final concentration of 100 $\mu$M, according to the information provided by the supplier, and stored in light-proof tubes at 4 $^{\circ}$C.

{\bf DNA sequences.} All the DNA sequences used in our study are listed in Supplementary Tables S1 and S2. The DNA sequences were designed to meet several requirements: (a) To balance the reaction rate, the GC content of all DNA strands was maintained at 50$\%$ $\sim$ 70$\%$; (b) We ensured the stability of the formed molecule; (c) To reduce the leakage of logic gates in the \text{DCMSubset} model, we avoided duplication of three consecutive bases at the same position in all toehold domains. Initially, the original sequences were obtained by using Nupack, and were then modified by hand.

{\bf Molecular Assembly.} All sequences that were assembled had to be checked to ensure that no secondary structures affecting strand displacement were generated. We found by NUPACK check that the generation rate of all logic gate secondary structures was not higher than 0.2$\%$ at 37$^{\circ}$C. Here, we used the NUPACK to obtain the minimum free energy structure (listed in Supplementary Table S3), where the relevant parameters required for simulation are shown in Supplementary Table S4. All $\Gamma$(x) complexes and $\Gamma'(x)$ complexes (listed in Supplementary Table S1) were assembled by mixing the corresponding single strands with equal molar concentrations and equal volumes. Moreover, all $D$(x) complexes (listed in Supplementary Table S1) were assembled by mixing the corresponding single strands using the concentration ratio  $\gamma$(x)$^*$: $output$ = 1.1 : 1. Depending on the buffer for dissolving $\gamma$(x)$^*$ and $output$ (which was sterilized ddH$_2$O), to ensure the same ion concentration of $\Gamma$(x) and $D$(x), we added an additional $3\times$ TAE containing 37.5 mM Mg$^{2+}$  to neutralize the ion concentration when assembling $D$(x), in addition to mixing $\gamma$(x)$^*$ and $output$. Further, $\gamma$(x)$^*$, $output$, and $3\times$ TAE/Mg$^{2+}$ were mixed with equal volume. We obtained the concentrations of the diluted DNA single strands using the instrument. We took two approaches to ensure the accuracy of the measured concentrations: (1) the solution was fully shaken and centrifuged before each measurement; and (2) three measurements were taken and averaged as the final concentration used.

Considering that length and number of single strands were different for the assembled computation and detection gates, we use two different annealing procedures to assemble logic gates. The computation gates $\Gamma$(x) and $\Gamma'(x)$ were first held at 95$^{\circ}$C for 3 minutes, then cooled to 60$^{\circ}$C at a rate of 1$^{\circ}$C/min, and then cooled to 4$^{\circ}$C at a rate of 0.33$^{\circ}$C/min. Finally, the computation gates were stored at 4 $^{\circ}$C. The detection gates were first held at 95$^{\circ}$C for 3 minutes, then cooled to 35$^{\circ}$C at a rate of 0.6$^{\circ}$C/min, and then cooled to 4$^{\circ}$C at a rate of 1$^{\circ}$C/min. Finally, the detection gates were stored in the dark at 4$^{\circ}$C.  These steps were performed on a PCR thermal cycler.

{\bf Agarose gel electrophoresis.} The DNA solutions were analyzed in a 5$\%$ Agarose gel in a 1$\times$ TAE buffer after running for 25 min at a constant power of 110 V using BG-subMINI and BG-Power 600, and imaged with the gel documentation system. 1$\times$ TAE was obtained by using ultrapure water to dilute 50$\times$ TAE. To prevent DNA samples from running out of the gel during electrophoresis, we used a DNA loading buffer (6$\times$) (2$\mu$l) to color the DNA samples (8$\mu$l) before performing electrophoresis. Also, we used DNA Marker A (25-500bp) (7$\mu$l) as a reference for the DNA samples (10$\mu$l).

The 5$\%$ agarose gel was prepared as follows: First, agarose (2.5 g powder) was dissolved in a 1$\times$TAE buffer (50 $ml$); then, it was heated until the powder was completely dissolved, and 4S Green Plus Nucleic Acid Stain (10 $\mu$l) was added to dissolve it fully; finally, it was poured into a membrane tool with a comb and cooled into a solid before being prepared for use.

{\bf Simulation and fluorescence kinetics experiments.} Simulation for dynamic analysis was completed by Visual DSD. The simulation duration was set to 600 s.

All fluorescence kinetic detection was performed using the QuantStudio 3 equipped with a 96-well fluorescence plate reader. The fluorescence detection procedure was as follows: First, in the Hold Stage, the temperature was decreased to 4 $^{\circ}$C by a rate of 1.6 $^{\circ}$C per second, and then the sample was quickly put into QuantStudio 3; second, in the PCR Stage, the temperature was increased to 23 $^{\circ}$C by a rate of 3 $^{\circ}$C per second; finally, the fluorescence intensity of the sample was measured every 3 min at 23 $^{\circ}$C.


\begin{thebibliography}{10}
\urlstyle{rm}
\expandafter\ifx\csname url\endcsname\relax
  \def\url#1{\texttt{#1}}\fi
\expandafter\ifx\csname urlprefix\endcsname\relax\def\urlprefix{URL }\fi
\expandafter\ifx\csname doiprefix\endcsname\relax\def\doiprefix{DOI: }\fi
\providecommand{\bibinfo}[2]{#2}
\providecommand{\eprint}[2][]{\url{#2}}

\bibitem{r1}
\bibinfo{author}{Xu, J.}, \bibinfo{author}{Qiang, X.}, \bibinfo{author}{Zhang,
  K.}, \bibinfo{author}{Zhang, C.} \& \bibinfo{author}{Yang, J.}
\newblock \bibinfo{journal}{\bibinfo{title}{A {DNA} computing model for the
  graph vertex coloring problem based on sa probe graph}}.
\newblock {\emph{\JournalTitle{Engineering}}} \textbf{\bibinfo{volume}{4}},
  \bibinfo{pages}{61--77} (\bibinfo{year}{2018}).

\bibitem{r2}
\bibinfo{author}{Tang, Z.} \emph{et~al.}
\newblock \bibinfo{journal}{\bibinfo{title}{Solving 0--1 integer programming
  problem based on {DNA} strand displacement reaction network}}.
\newblock {\emph{\JournalTitle{ACS Synthetic Biology}}}
  \textbf{\bibinfo{volume}{10}}, \bibinfo{pages}{2318--2330}
  (\bibinfo{year}{2021}).

\bibitem{r3}
\bibinfo{author}{Benenson, Y.}
\newblock \bibinfo{journal}{\bibinfo{title}{{DNA} computes a square root}}.
\newblock {\emph{\JournalTitle{Nature Nanotechnol}}}
  \textbf{\bibinfo{volume}{6}}, \bibinfo{pages}{465--467}
  (\bibinfo{year}{2011}).

\bibitem{r4}
\bibinfo{author}{Xu, J.}, \bibinfo{author}{Chen, C.} \& \bibinfo{author}{Shi,
  X.}
\newblock \bibinfo{journal}{\bibinfo{title}{Graph computation using algorithmic
  self-assembly of {DNA} molecules}}.
\newblock {\emph{\JournalTitle{ACS Synthetic Biology}}}
  \textbf{\bibinfo{volume}{11}}, \bibinfo{pages}{2456--2463}
  (\bibinfo{year}{2022}).

\bibitem{r5}
\bibinfo{author}{Qian, L.}, \bibinfo{author}{Winfree, E.} \&
  \bibinfo{author}{Bruck, J.}
\newblock \bibinfo{journal}{\bibinfo{title}{Neural network computation with
  {DNA} strand displacement cascades}}.
\newblock {\emph{\JournalTitle{Nature}}} \textbf{\bibinfo{volume}{475}},
  \bibinfo{pages}{368--372} (\bibinfo{year}{2011}).

\bibitem{r6}
\bibinfo{author}{Qian, L.} \& \bibinfo{author}{Winfree, E.}
\newblock \bibinfo{journal}{\bibinfo{title}{Scaling up digital circuit
  computation with {DNA} strand displacement cascades}}.
\newblock {\emph{\JournalTitle{Science}}} \textbf{\bibinfo{volume}{332}},
  \bibinfo{pages}{1196--1201} (\bibinfo{year}{2011}).

\bibitem{r7}
\bibinfo{author}{Jung, J.~K.}, \bibinfo{author}{Archuleta, C.~M.},
  \bibinfo{author}{Alam, K.~K.} \& \bibinfo{author}{Lucks, J.~B.}
\newblock \bibinfo{journal}{\bibinfo{title}{Programming cell-free biosensors
  with {DNA} strand displacement circuits}}.
\newblock {\emph{\JournalTitle{Nature chem biol}}}
  \textbf{\bibinfo{volume}{18}}, \bibinfo{pages}{385--393}
  (\bibinfo{year}{2022}).

\bibitem{r8}
\bibinfo{author}{Jung, J.~K.} \emph{et~al.}
\newblock \bibinfo{journal}{\bibinfo{title}{Cell-free biosensors for rapid
  detection of water contaminants}}.
\newblock {\emph{\JournalTitle{Nature biotechnol}}}
  \textbf{\bibinfo{volume}{38}}, \bibinfo{pages}{1451--1459}
  (\bibinfo{year}{2020}).

\bibitem{r9}
\bibinfo{author}{Adleman, L.~M.}
\newblock \bibinfo{journal}{\bibinfo{title}{Molecular computation of solutions
  to combinatorial problems}}.
\newblock {\emph{\JournalTitle{Science}}} \textbf{\bibinfo{volume}{266}},
  \bibinfo{pages}{1021--1024} (\bibinfo{year}{1994}).

\bibitem{r10}
\bibinfo{author}{Reif, J.~H.}
\newblock \bibinfo{title}{Parallel molecular computation}.
\newblock In \emph{\bibinfo{booktitle}{Proceedings of the seventh annual ACM
  symposium on Parallel algorithms and architectures}},
  \bibinfo{pages}{213--223} (\bibinfo{year}{1995}).

\bibitem{r11}
\bibinfo{author}{Lo, Y.-M.}, \bibinfo{author}{Yiu, K.} \&
  \bibinfo{author}{Wong, S.}
\newblock \bibinfo{journal}{\bibinfo{title}{On the potential of molecular
  computing}}.
\newblock {\emph{\JournalTitle{Science}}} \textbf{\bibinfo{volume}{268}},
  \bibinfo{pages}{481--482} (\bibinfo{year}{1995}).

\bibitem{r12}
\bibinfo{author}{Lipton, R.~J.}
\newblock \bibinfo{journal}{\bibinfo{title}{{DNA} solution of hard
  computational problems}}.
\newblock {\emph{\JournalTitle{Science}}} \textbf{\bibinfo{volume}{268}},
  \bibinfo{pages}{542--545} (\bibinfo{year}{1995}).

\bibitem{r13}
\bibinfo{author}{Roweis, S.} \emph{et~al.}
\newblock \bibinfo{journal}{\bibinfo{title}{A sticker-based model for {DNA}
  computation}}.
\newblock {\emph{\JournalTitle{Journal of Computational Biology}}}
  \textbf{\bibinfo{volume}{5}}, \bibinfo{pages}{615--629}
  (\bibinfo{year}{1998}).

\bibitem{r14}
\bibinfo{author}{Zimmermann, K.-H.}
\newblock \bibinfo{journal}{\bibinfo{title}{Efficient dna sticker algorithms
  for {NP}-complete graph problems}}.
\newblock {\emph{\JournalTitle{Computer Physics Communications}}}
  \textbf{\bibinfo{volume}{144}}, \bibinfo{pages}{297--309}
  (\bibinfo{year}{2002}).

\bibitem{r15}
\bibinfo{author}{Braich, R.~S.}, \bibinfo{author}{Chelyapov, N.},
  \bibinfo{author}{Johnson, C.}, \bibinfo{author}{Rothemund, P.~W.} \&
  \bibinfo{author}{Adleman, L.}
\newblock \bibinfo{journal}{\bibinfo{title}{Solution of a 20-variable 3-{SAT}
  problem on a dna computer}}.
\newblock {\emph{\JournalTitle{Science}}} \textbf{\bibinfo{volume}{296}},
  \bibinfo{pages}{499--502} (\bibinfo{year}{2002}).

\bibitem{r16}
\bibinfo{author}{Yurke, B.}, \bibinfo{author}{Turberfield, A.~J.},
  \bibinfo{author}{Mills, A.~P.}, \bibinfo{author}{Simmel, F.~C.} \&
  \bibinfo{author}{Neumann, J.~L.}
\newblock \bibinfo{journal}{\bibinfo{title}{A {DNA}-fuelled molecular machine
  made of dna}}.
\newblock {\emph{\JournalTitle{Nature}}} \textbf{\bibinfo{volume}{406}},
  \bibinfo{pages}{605--608} (\bibinfo{year}{2000}).

\bibitem{r17}
\bibinfo{author}{Liu, C.} \emph{et~al.}
\newblock \bibinfo{journal}{\bibinfo{title}{Cross-inhibitor: a time-sensitive
  molecular circuit based on {DNA} strand displacement}}.
\newblock {\emph{\JournalTitle{Nucleic Acids Research}}}
  \textbf{\bibinfo{volume}{48}}, \bibinfo{pages}{10691--10701}
  (\bibinfo{year}{2020}).

\bibitem{r18}
\bibinfo{author}{Zhang, D.~Y.} \& \bibinfo{author}{Winfree, E.}
\newblock \bibinfo{journal}{\bibinfo{title}{Control of dna strand displacement
  kinetics using toehold exchange}}.
\newblock {\emph{\JournalTitle{Journal of the American Chemical Society}}}
  \textbf{\bibinfo{volume}{131}}, \bibinfo{pages}{17303--17314}
  (\bibinfo{year}{2009}).

\bibitem{r19}
\bibinfo{author}{Machinek, R.~R.}, \bibinfo{author}{Ouldridge, T.~E.},
  \bibinfo{author}{Haley, N.~E.}, \bibinfo{author}{Bath, J.} \&
  \bibinfo{author}{Turberfield, A.~J.}
\newblock \bibinfo{journal}{\bibinfo{title}{Programmable energy landscapes for
  kinetic control of {DNA} strand displacement}}.
\newblock {\emph{\JournalTitle{Nature Communications}}}
  \textbf{\bibinfo{volume}{5}}, \bibinfo{pages}{1--9} (\bibinfo{year}{2014}).

\bibitem{r20}
\bibinfo{author}{Wang, F.} \emph{et~al.}
\newblock \bibinfo{journal}{\bibinfo{title}{Implementing digital computing with
  dna-based switching circuits}}.
\newblock {\emph{\JournalTitle{Nature Communications}}}
  \textbf{\bibinfo{volume}{11}}, \bibinfo{pages}{1--8} (\bibinfo{year}{2020}).

\bibitem{r21}
\bibinfo{author}{Zhu, E.}, \bibinfo{author}{Luo, X.}, \bibinfo{author}{Liu, C.}
  \& \bibinfo{author}{Chen, C.}
\newblock \bibinfo{journal}{\bibinfo{title}{An operational {DNA} strand
  displacement encryption approach}}.
\newblock {\emph{\JournalTitle{Nanomaterials}}} \textbf{\bibinfo{volume}{12}},
  \bibinfo{pages}{877} (\bibinfo{year}{2022}).

\bibitem{r22}
\bibinfo{author}{Tang, Z.}, \bibinfo{author}{Yin, Z.}, \bibinfo{author}{Yang,
  J.}, \bibinfo{author}{Sun, X.} \& \bibinfo{author}{Cui, J.}
\newblock \bibinfo{title}{The circular {DNA} model of 0--1 programming problem
  based on dna strand displacement}.
\newblock In \emph{\bibinfo{booktitle}{2018 14th International Conference on
  Natural Computation, Fuzzy Systems and Knowledge Discovery (ICNC-FSKD)}},
  \bibinfo{pages}{173--177} (\bibinfo{organization}{IEEE},
  \bibinfo{year}{2018}).

\bibitem{r23}
\bibinfo{author}{Yang, J.}, \bibinfo{author}{Pang, X.}, \bibinfo{author}{Tang,
  Z.}, \bibinfo{author}{Yang, X.} \& \bibinfo{author}{Liu, C.}
\newblock \bibinfo{title}{{DNA} strand displacement computing model for the sat
  problem}.
\newblock In \emph{\bibinfo{booktitle}{Journal of Physics: Conference Series}},
  vol. \bibinfo{volume}{2026}, \bibinfo{pages}{012040}
  (\bibinfo{organization}{IOP Publishing}, \bibinfo{year}{2021}).

\bibitem{r24}
\bibinfo{author}{Zhang, X.}, \bibinfo{author}{Zhang, W.},
  \bibinfo{author}{Zhao, T.}, \bibinfo{author}{Wang, Y.} \&
  \bibinfo{author}{Cui, G.}
\newblock \bibinfo{journal}{\bibinfo{title}{Design of logic circuits based on
  combinatorial displacement of {DNA} strands}}.
\newblock {\emph{\JournalTitle{Journal of Computational and Theoretical
  Nanoscience}}} \textbf{\bibinfo{volume}{12}}, \bibinfo{pages}{1161--1164}
  (\bibinfo{year}{2015}).

\bibitem{r25}
\bibinfo{author}{Yang, C.} \emph{et~al.}
\newblock \bibinfo{journal}{\bibinfo{title}{A versatile {DNA}-supramolecule
  logic platform for multifunctional information processing}}.
\newblock {\emph{\JournalTitle{NPG Asia Materials}}}
  \textbf{\bibinfo{volume}{10}}, \bibinfo{pages}{497--508}
  (\bibinfo{year}{2018}).

\bibitem{r26}
\bibinfo{author}{Woods, D.} \emph{et~al.}
\newblock \bibinfo{journal}{\bibinfo{title}{Diverse and robust molecular
  algorithms using reprogrammable {DNA} self-assembly}}.
\newblock {\emph{\JournalTitle{Nature}}} \textbf{\bibinfo{volume}{567}},
  \bibinfo{pages}{366--372} (\bibinfo{year}{2019}).

\bibitem{r27}
\bibinfo{author}{Currin, A.} \emph{et~al.}
\newblock \bibinfo{journal}{\bibinfo{title}{Computing exponentially faster:
  implementing a non-deterministic universal turing machine using {DNA}}}.
\newblock {\emph{\JournalTitle{Journal of the R Society Interface}}}
  \textbf{\bibinfo{volume}{14}}, \bibinfo{pages}{20160990}
  (\bibinfo{year}{2017}).

\bibitem{r28}
\bibinfo{author}{Lee, W.}, \bibinfo{author}{Yu, M.}, \bibinfo{author}{Lim, D.},
  \bibinfo{author}{Kang, T.} \& \bibinfo{author}{Song, Y.}
\newblock \bibinfo{journal}{\bibinfo{title}{Programmable {DNA}-based boolean
  logic microfluidic processing unit}}.
\newblock {\emph{\JournalTitle{ACS Nano}}} \textbf{\bibinfo{volume}{15}},
  \bibinfo{pages}{11644--11654} (\bibinfo{year}{2021}).

\bibitem{r29}
\bibinfo{author}{Cook, M.} \emph{et~al.}
\newblock \bibinfo{journal}{\bibinfo{title}{Universality in elementary cellular
  automata}}.
\newblock {\emph{\JournalTitle{Complex Systems}}}
  \textbf{\bibinfo{volume}{15}}, \bibinfo{pages}{1--40} (\bibinfo{year}{2004}).

\bibitem{r30}
\bibinfo{author}{Margulies, D.}, \bibinfo{author}{Melman, G.} \&
  \bibinfo{author}{Shanzer, A.}
\newblock \bibinfo{journal}{\bibinfo{title}{A molecular full-adder and
  full-subtractor, an additional step toward a moleculator}}.
\newblock {\emph{\JournalTitle{Journal of the American Chemical Society}}}
  \textbf{\bibinfo{volume}{128}}, \bibinfo{pages}{4865--4871}
  (\bibinfo{year}{2006}).

\bibitem{r31}
\bibinfo{author}{Zhou, C.}, \bibinfo{author}{Geng, H.}, \bibinfo{author}{Wang,
  P.} \& \bibinfo{author}{Guo, C.}
\newblock \bibinfo{journal}{\bibinfo{title}{Programmable {DNA}
  nanoindicator-based platform for large-scale square root logic
  biocomputing}}.
\newblock {\emph{\JournalTitle{Small}}} \textbf{\bibinfo{volume}{15}},
  \bibinfo{pages}{1903489} (\bibinfo{year}{2019}).

\bibitem{r32}
\bibinfo{author}{Xie, T.} \emph{et~al.}
\newblock \bibinfo{journal}{\bibinfo{title}{{DNA} circuits compatible encoder
  and demultiplexer based on a single biomolecular platform with dna strands as
  outputs}}.
\newblock {\emph{\JournalTitle{Nucleic Acids Research}}}
  \textbf{\bibinfo{volume}{50}}, \bibinfo{pages}{8431--8440}
  (\bibinfo{year}{2022}).

\bibitem{r33}
\bibinfo{author}{Shah, S.} \emph{et~al.}
\newblock \bibinfo{journal}{\bibinfo{title}{Using strand displacing polymerase
  to program chemical reaction networks}}.
\newblock {\emph{\JournalTitle{Journal of the American Chemical Society}}}
  \textbf{\bibinfo{volume}{142}}, \bibinfo{pages}{9587--9593}
  (\bibinfo{year}{2020}).

\bibitem{r34}
\bibinfo{author}{Reid, M.~S.}, \bibinfo{author}{Le, X.~C.} \&
  \bibinfo{author}{Zhang, H.}
\newblock \bibinfo{journal}{\bibinfo{title}{Exponential isothermal
  amplification of nucleic acids and assays for proteins, cells, small
  molecules, and enzyme activities: an expar example}}.
\newblock {\emph{\JournalTitle{Angew. Chem. International. Edit}}}
  \textbf{\bibinfo{volume}{57}}, \bibinfo{pages}{11856--11866}
  (\bibinfo{year}{2018}).

\bibitem{r35}
\bibinfo{author}{Kang, D.} \emph{et~al.}
\newblock \bibinfo{journal}{\bibinfo{title}{{DNA} biomolecular-electronic
  encoder and decoder devices constructed by multiplex biosensors}}.
\newblock {\emph{\JournalTitle{NPG Asia Materials}}}
  \textbf{\bibinfo{volume}{4}}, \bibinfo{pages}{e1--e1} (\bibinfo{year}{2012}).

\bibitem{r36}
\bibinfo{author}{de~Luis, B.} \emph{et~al.}
\newblock \bibinfo{journal}{\bibinfo{title}{A 1-to-2 demultiplexer hybrid
  nanocarrier for cargo delivery and activation}}.
\newblock {\emph{\JournalTitle{Chemical Communications}}}
  \textbf{\bibinfo{volume}{56}}, \bibinfo{pages}{9974--9977}
  (\bibinfo{year}{2020}).

\bibitem{r37}
\bibinfo{author}{Thomas, P.} \emph{et~al.}
\newblock \bibinfo{journal}{\bibinfo{title}{1-to-4 analog demultiplexer with up
  to 128 gs/s for interleaving of bandwidth-limited digitizers in wireline and
  optical receivers}}.
\newblock {\emph{\JournalTitle{IEEE. Journal of Solid-State Circuits}}}
  \textbf{\bibinfo{volume}{56}}, \bibinfo{pages}{2611--2623}
  (\bibinfo{year}{2021}).

\bibitem{r38}
\bibinfo{author}{Reed, B.}
\newblock \bibinfo{journal}{\bibinfo{title}{Paths, stars and the number
  three}}.
\newblock {\emph{\JournalTitle{Combinatorics, Probability and Computing}}}
  \textbf{\bibinfo{volume}{5}}, \bibinfo{pages}{277--295}
  (\bibinfo{year}{1996}).

\bibitem{r39}
\bibinfo{author}{Lov{\'a}sz, L.}
\newblock \bibinfo{journal}{\bibinfo{title}{On the shannon capacity of a
  graph}}.
\newblock {\emph{\JournalTitle{IEEE. Trans. Information Theory}}}
  \textbf{\bibinfo{volume}{25}}, \bibinfo{pages}{1--7} (\bibinfo{year}{1979}).

\bibitem{r40}
\bibinfo{author}{Brooks, R.~L.}
\newblock \bibinfo{journal}{\bibinfo{title}{On colouring the nodes of a
  network}}.
\newblock {\emph{\JournalTitle{Mathematical Proceedings of the Cambridge
  Philosophical Society}}} \textbf{\bibinfo{volume}{37}},
  \bibinfo{pages}{194--197} (\bibinfo{year}{1941}).

\bibitem{r41}
\bibinfo{author}{Gallai, T.}
\newblock \bibinfo{journal}{\bibinfo{title}{{\"{U}}ber extreme punkt- und
  kantenmengen}}.
\newblock {\emph{\JournalTitle{Annales Universitatis Scientiarium
  Budapestinensis de Rolando E{\"{o}}tv{\"{o}}s Nominatae, Sectio
  Mathematica}}} \textbf{\bibinfo{volume}{2}}, \bibinfo{pages}{133--138}
  (\bibinfo{year}{1959}).

\end{thebibliography}

\noindent\textbf{Acknowledgements}\\
This work was supported in part by  National Natural Science Foundation of China  under Grant 61872101; in part by Natural Science Foundation of Guangdong Province of China under Grant 2021A1515011940.\\

\noindent\textbf{Author contributions}\\
Conceptualization, E.Z. and X.L.; Investigation, E.Z., X.L. and C.L.; Methodology, E.Z., X.L. and X.S.; Supervision, J.X. and X.S.; Writing-original draft, E.Z. and  X.L.; Writing-review and editing, E.Z. and C.L. All authors have read and agreed to the published version of the manuscript.\\

\noindent\textbf{Competing interests}\\
The authors declare no conflict of interest.\\

\noindent\textbf{Supplementary information}
Supplementary information is available for this paper at \href{http://www.nature.com/authors/policies/competing.html}{here}.

\end{multicols}
\end{document}